\documentclass[12pt,intlim,righttag]{article}

\usepackage{amscd,amsfonts,amssymb,amsmath,,amsthm,latexsym,array,hhline}

\newcommand{\sg}{\sigma}

\newcommand{\al}{\alpha}
\newcommand{\gm}{\gamma}

\newcommand{\vp}{\varphi}

\newcommand{\og}{\omega}

\newcommand{\be}{\begin}
\newcommand{\ee}{\end}
\newcommand{\lbl}{\label}

\newcommand\beq{\begin{equation}}
\newcommand\eeq{\end{equation}}

\theoremstyle{Theorem}

\theoremstyle{corollary}

\theoremstyle{remark}

\theoremstyle{definition}

\begin{document}

\title{Robust mean-variance hedging in the single period model}

\author{R. Tevzadze $^{1),2)}$ and T. Uzunashvili $^{1)}$}
\date{~}
\maketitle

\begin{center}
{$^{1)}$ Georgian--American University, Business School, 3,
Alleyway II,
\newline
Chavchavadze Ave. 17\,a,
\newline $^{2)}$ Institute of Cybernetics,  5 Euli str., 0186, Tbilisi,
Georgia
\newline
(e-mail: rtevzadze@gmail.com) }
\end{center}

\numberwithin{equation}{section}

\begin{abstract}

We give an explicit solution of robust mean-variance hedging
problem in the  single period model for some type of contingent
claims. The alternative approach is also considered.

\bigskip

\noindent {\bf Key words and phrases}:{The min-max problem,
mean-variance hedging, maximum principle, robust optimization.}

\noindent {\bf  Mathematics Subject Classification (2000)}:
60H30,90C47.

\end{abstract}

\

\section{Introduction}

\

The study of mean variance hedging problem was initiated by H.
F\"ollmer and D. Sondermann \cite{F-S} and the solution of this
problem for multiperiod model was given by H. F\"ollmer and M.
Schweizer \cite{F-Sch}. In this paper we investigate the single
period mean variance hedging problem of contingent claims in
incomplete markets, when parameters of asset prices are not known
with certainty. Usually such parameters may be appreciate rate (or
drift) and volatility coefficients. In such models it is desirable
to choose an optimal  portfolio for the worst case of parameters.
Such type problem one calls the robust hedging problem.

The numerous of publications are concerned to the case when one of
these parameters is known exactly. In the case of unknown drift
coefficient the existence of saddle point of corresponding minimax
problem has been established and characterization of the optimal
strategy has been obtained (see
\cite{CvKa},\cite{herher},\cite{gun}). For the case of unknown
volatility coefficients the construction of hedging strategy  were
given in the works \cite{ave}, \cite{ah}, \cite{ahn}, \cite{lato}.

The most difficult case is to characterize the optimal strategy of
minimax (or maximin) problem under uncertainty of both drift and
volatility terms.
 Talay and Zheng \cite{TaZ} applied the PDE-based approach to
 the maximin problem in the continuous time model and
 characterized
 the value as a viscosity solution of corresponding
 Bellman-Isaacs equation. However for robust hedging it is more
 convenient to consider the minimax problem.
Such type of problem was studied  for the single period model of
financial market by Pinar \cite{pin}, who  consider the
computational scheme to find the optimal strategy and optimal
initial capital.

 The purpose
of the present paper is to investigate the robust mean-variance
hedging problem in the one-step  model, when drift and volatility
of the asset are not known exactly. We consider the minimax
problem and construct the optimal strategy for some type of
contingent claims. Our approach is twofold. The main approach we
develop is the randomization of the parameters and change the
minimax problem by maximin one. This approach successfully works
in the one period model and preliminary results show that it will
be productive in multi-period and continuous time models. The
other way is to perform maximization and minimization directly as
they are given and  describe the solution based on result of
\cite{dem}. This way we call the alternative.

The paper is organized as follows. In section 2 we describe the
market model and give a setting of the problem. Using
randomization of parameters we argue the existence of saddle
point. Further we construct explicit solution of obtained maximin
problem. In examples 1-3 are considered the particular cases when
the optimal strategy is expressed in a simple form.   In section 3
we give an alternative approach to the minimax problem based on
the result of \cite{dem}.

\

\section{The main results}

\

We consider a financial market model with two assets. Let
$(S_t,\eta_t),\;t=0,1$ be the price of assets. We suppose that
\beq S_1=S_0+\mu+\sg w,\;\eta_1=\beta+\delta\bar w,\eeq where
$w,\bar w$ is random pair with $Ew=E\bar w=0,\;Dw=D\bar
w=1,\;Cov(w,\bar w)\neq 0$ and $\mu,\sg,\beta,\delta$ are
constants. We suppose also that the appreciate rate $\mu$ and
volatility $\sg$ of the asset price $S_t$ are misspecified but
stay in rectangle of uncertainty, i.e.
$$(\mu,\sg)\in D=[{\mu_-},\mu_+]\times[\sg_-,\sg_+].$$
Let $\beta,\delta$ be known exactly. We denote by $\pi$ the number
of stocks $S$ bought at time $t=0$ and by $x_0=\pi S_0$ the
initial capital. The  wealth at time $t=1$ is
$$X_1=x_0+\pi(S_1-S_0)=x_0+\pi\mu+\pi\sg w.$$ The contingent claim $H(\eta)$ we
assume depends on the asset $\eta$, which cannot be traded
directly. The robust mean-variance hedging problem is \beq
\min_\pi\max_{\mu,\sg}E|H-x_0-\pi\mu-\pi\sg w|^2.\eeq

Let \beq H-x_0=h_0+h_1w+H^\bot\eeq be the decomposition of $H-x_0$
with $h_0=E(H-x_0),\;h_1=EwH,\;EwH^\bot=0$. Then  the problem can
be rewritten as
\begin{equation}\lbl{mnmx}
\min_{\pi\in R}\max_{(\mu,\sg)\in D}F(\pi,\mu,\sg),
\end{equation}
where
\begin{equation}
F(\pi,\mu,\sg)=(h_0-\pi\mu)^2+(h_1-\pi\sg)^2.
\end{equation}
The function $F(\pi,\cdot)$ can be continued on the space of
probability measures on $D$ as
\begin{equation}
F(\pi,\nu)=\int_D((h_0-\pi\mu)^2+(h_1-\pi\sg)^2)\nu(d\mu
d\sg),\;\;\text{for measure}\; \nu\; \text{on}\; D
\end{equation}
Hence we get
\begin{align}
F(\pi,\nu)=\int_D(\mu^2+\sg^2)\nu(d\mu
d\sg)\left(\pi-\frac{\int_D(h_0\mu+h_1\sg)\nu(d\mu
d\sg)}{\int_D(\mu^2+\sg^2)\nu(d\mu d\sg)}\right)^2
\\
+h_0^2+h_1^2-\frac{(\int_D(h_0\mu+h_1\sg)\nu(d\mu
d\sg))^2}{\int_D(\mu^2+\sg^2)\nu(d\mu d\sg)}
\end{align}
and
\begin{equation}
\min_{\pi\in
R}F(\pi,\nu)=h_0^2+h_1^2-\frac{(\int_D(h_0\mu+h_1\sg)\nu(d\mu
d\sg))^2}{\int_D(\mu^2+\sg^2)\nu(d\mu d\sg)}
\end{equation}
\begin{equation}\lbl{opst}
\pi^*=\frac{\int_D(h_0\mu+h_1\sg)\nu(d\mu
d\sg)}{\int_D(\mu^2+\sg^2)\nu(d\mu d\sg)}
\end{equation}
Since $F$ is strictly convex in $\pi$ by the Theorem Neumann at
al. (see Theorem lX.4.1 of \cite{var}) there exists a saddle point
$(\pi^*,\nu^*)$, i.e.
\begin{equation}
F(\pi^*,\nu)\le F(\pi^*,\nu^*)\le F(\pi,\nu^*).
\end{equation}
Since $\max_\nu F(\pi,\nu)=\max_{\mu,\sg}F(\pi,\mu,\sg)$ then we
obtain
\begin{equation}
\min_\pi\max_{(\mu,\sg)}F(\pi,\mu,\sg)=\min_{\pi}\max_{\nu}
F(\pi,\nu)=\max_{\nu}\min_\pi F(\pi,\nu).
\end{equation}
Each pair of random variables $(\mu,\sg)$ with the distribution
$\nu$ may be realized on the probability space $([0.1],{\cal
B},P(d\omega)=d\omega)$ where ${\cal B}$ is the Borel
$\sg-$algebra on $[0,1]$ and $d\omega$ the Lebesgue measure (see
Proposition 26.6 of \cite{part}). Hence the minimization problem
\begin{equation}\lbl{mop}
\min_{\nu}\frac{(\int_D(h_0\mu+h_1\sg)\nu(d\mu
d\sg))^2}{\int_D(\mu^2+\sg^2)\nu(d\mu d\sg)}
\end{equation}
can be written as
\begin{equation}\lbl{mop1}
\min_{(\mu(\og),\sg(\og))\in D
}\frac{(\int_0^1(h_0\mu(\og)+h_1\sg(\og))d\og)^2}{\int_0^1(\mu^2(\og)+\sg^2(\og))d\og}.
\end{equation}
To solve this problem we consider the deterministic control
problem
\begin{align}\lbl{mop2}
\max_{(\mu(\og),\sg(\og))\in D
}{\int_0^1(\mu^2(\og)+\sg^2(\og))d\og},\\
\frac{dx(\og)}{d\og}=\mu(\og)\;\;\frac{dy(\og)}{d\og}=\sg(\og),\\
\lbl{cxy}x(0)=0,y(0)=0,\;\;x(1)=x,y(1)=y.
\end{align}
\be{lem} The solution of the problem (\ref{mop2}) is of the form
\begin{equation}\lbl{ms}
\mu^*(\og)=\mu_-\chi_A(\og)+\mu_+\chi_{A^c}(\og),\;\;\sg^*(\og)=\sg_-\chi_B(\og)+\sg_+\chi_{B^c}(\og),
\end{equation}
with
\begin{equation}\lbl{xy}
P(A)=\frac{x-\mu_-}{\mu_+-\mu_-},\;\;P(B)=\frac{y-\sg_-}{\sg_+-\sg_-}
\end{equation}
and the maximal value is $2x\mu_M+2y\sg_M-\mu_-\mu_+-\sg_-\sg_+ $,
where $\mu_M=\frac{\mu_++\mu_-}{2},\;\sg_M=\frac{\sg_++\sg_-}{2}$.
 \ee{lem}
{\it Proof}. By the maximum principle (see \cite{var}) we have
$$
\mu^*=\arg\max_{\mu_-\le\mu\le\mu_+}{(\mu^2+p\mu)},\;\sg^*=\arg\max_{\sg_-\le\sg\le\sg_+}{(\sg^2+q\sg)},
$$
where $p,q$ are some constants maintaining the conditions
(\ref{cxy}). Hence the solution of the problem (\ref{mop2}) is of
the form (\ref{ms}). The relations
$$\int_0^1\mu^*(\og)d\og=x,\;\;\int_0^1\sg^*(\og)d\og=y$$
uniquely determines the probabilities $P(A),P(B)$ by (\ref{xy})
and
$$\int_0^1(\mu^{*2}(\og)+\sg^{*2}(\og))d\og=2x\mu_M+2y\sg_M-\mu_-\mu_+-\sg_-\sg_+.$$

\be{cor}
\begin{equation}
\min_{(\mu(\og),\sg(\og))\in D
}\frac{(\int_0^1(h_0\mu(\og)+h_1\sg(\og))d\og)^2}{\int_0^1(\mu^2(\og)+\sg^2(\og))d\og}=\min_{(x,y)\in
D} \frac{(h_0x+h_1y)^2}{2\mu_Mx+2\sg_My-\mu_-\mu_+-\sg_-\sg_+}.
\end{equation}
\ee{cor}

To characterize the minimum point of function \be{equation}
\psi(x,y)=\frac{(h_0x+h_1y)^2}{2\mu_Mx+2\sg_My-\mu_-\mu_+-\sg_-\sg_+}
\ee{equation} we use the following lemma.

\be{lem} The solution of the system \be{equation}
\frac{\partial\psi}{\partial
x}(x,y)=0,\;\;\frac{\partial\psi}{\partial y}(x,y)=0 \ee{equation}
satisfies the equation $h_0x+h_1y=0$. \ee{lem}

{\it Proof}. It is easy to see that
\begin{align*}
\frac{\partial\psi}{\partial x}(x,y)
=2(h_0x+h_1y)\\
\times\frac{h_0(2\mu_Mx+2\sg_My-\mu_-\mu_+-\sg_-\sg_+)-h_0\mu_Mx-h_1\mu_My}
{(2\mu_Mx+2\sg_My-\mu_-\mu_+-\sg_-\sg_+)^2},\\
\frac{\partial\psi}{\partial y}(x,y)
=2(h_0x+h_1y)\\
\times\frac{h_1(2\mu_Mx+2\sg_My-\mu_-\mu_+-\sg_-\sg_+)-h_0\sg_Mx-h_1\sg_My}
{(2\mu_Mx+2\sg_My-\mu_-\mu_+-\sg_-\sg_+)^2}. \end{align*} Solving
the system we obtain that either $h_0x+h_1y=0$ or
\begin{align*}
h_0\mu_Mx+(2h_0\sg_M-h_1\mu_M)y=
h_0(\mu_+\mu_++\sg_-\sg_+),\\
h_1\sg_My+(2h_1\mu_M-h_0\sg_M)x=
h_1(\mu_+\mu_++\sg_-\sg_+).\end{align*} The latter system admits
the unique solution
\begin{align*}
x=\frac{h_1}{2}\frac{\mu_-\mu_++\sg_-\sg_+}{h_1\mu_M-h_0\sg_M},\\
y=-\frac{h_0}{2}\frac{\mu_-\mu_++\sg_-\sg_+}{h_1\mu_M-h_0\sg_M},\end{align*}
which also satisfies the equation $h_0x+h_1y=0$.

\be{cor}\lbl{p1} The minimum of $\psi(x,y)\in D$ is achieved
either on the line $h_0x+h_1y=0$ or on the boundary of $D$.
\ee{cor}

\be{lem} If there exists the pair $(\bar x,\bar y)$ such that
$h_0\bar x+h_1\bar y=0$, then
$$
\max_\nu\min_\pi F(\pi,\nu)=\min_{(x,y)\in D}\psi(x,y)= \psi(\bar
x,\bar y)
$$
and $\pi^*=0$.
 \ee{lem}
{\it Proof}. It is sufficient to take $(\mu^*,\sg^*)=(\bar x ,\bar
y)$ and to use (\ref{opst}). \qed

From now on we assume that $h_0x+h_1y\neq 0$ for all $(x,y)\in D$.
For certainty we suppose that $h_0x+h_1y>0$. The case
$h_0x+h_1y<0$ can be considered analogously.

The boundary $\partial D$ of rectangle $D$ consists from the sides
$B_{--},B_{-+},B_{+-},B_{++},$ where
\begin{align*}B_{--}=\{(x,y):x=\mu_-,\;\sg_-\le
y\le\sg_+\},\\
B_{-+}=\{(x,y):x=\mu_+,\;\sg_-\le y\le\sg_+\},\\
B_{+-}=\{(x,y):y=\sg_-,\;\mu_-\le
x\le\mu_+\},\\
B_{++}=\{(x,y):y=\sg_+,\;\mu_-\le x\le\mu_+\}. \end{align*}
Obviously that functions defined on the sides
\begin{align*}\vp_{-a}(t)=\psi(\mu_a,\sg_-+t(\sg_+-\sg_-)),\;on\;\;B_{-a},\;\;a=-,+,\\
\vp_{-b}(t)=\psi(\mu_-+t(\mu_+-\mu_-),\sg_b),\;on\;\;B_{+b},\;\;b=-,+,\end{align*}
coincide with functions of the Appendix. It is easy  to show that
the $t_{ab}=\arg\min\vp_{ab}(t),\;a=-,+,\;b=+,-$ can be computed
as (see Appendix)

\begin{equation}\lbl{tab}
t_{ab}=
  \begin{cases}
    1, & \text{if}\;\;1\le \al_{ab}\;\;\text{or}\;\;1\le 2\beta_{ab}-\al_{ab}, \\
    0, & \text{if} \;\; \beta_{ab}\le \al_{ab}\le0 \;\;\text{or}\;\;\beta_{ab}<2\beta_{ab}-\al_{ab}\le 0\\
    \al_{ab}, & \text{if} \;\; 0<\al_{ab}<1 \\
    2\beta_{ab}-\al_{ab}, & \text{if}\;\;0<2\beta_{ab}-\al_{ab}<1.
  \end{cases}
\end{equation}
Hence we have

\be{prop}\lbl{p1} Let $h_0x+h_1y>0$ for all $(x,y)\in D$. Then
$$
\min_{(x,y)\in D}\psi(x,y)=\min_{a=\pm,b=\pm}\vp_{ab}(t_{ab}).
$$
Moreover for $(x^*,y^*)=\arg\min_{(x,y)\in D}\psi(x,y)$ we have
$(x^*,y^*)\in B_{a^*b^*}$, where
$a^*b^*=\arg\min_{ab}\vp_{ab}(t_{ab})$ and  $t^*= t_{a^*b^*}$ is
the distance from the end of the side to $(x^*,y^*)$ defined by
(\ref{tab}) . \ee{prop}

\be{prop} Let $h_0x+h_1y> 0$ for all $(x,y)\in D$. Then the
solution of the optimization problem (\ref{mop1}) is of the form
 \be{equation}
(\mu^*,\sg^*)=\be{cases}(\mu_-,\sg_-\chi_B+\sg_+\chi_{B^c}),\;\text{if}\;(x^*,y^*)\in
B_{--},\\(\mu_+,\sg_-\chi_B+\sg_+\chi_{B^c}),
\;\text{if}\;(x^*,y^*)\in B_{-+},\\
(\mu_-\chi_A+\mu_+\chi_{A^c},\sg_-),\;\text{if}\;(x^*,y^*)\in
B_{+-},\\(\mu_-\chi_A+\mu_+\chi_{A^c},\sg_+),\;\text{if}\;(x^*,y^*)\in
B_{++}. \ee{cases}\ee{equation} \ee{prop}

{\it Proof}. Let $(x^*,y^*)$ be the minimum point of $\psi(x,y)$.
By Proposition (\ref{p1}) $(x^*,y^*)$ belongs on some side. Hence
the pair $(\mu^*,\sg^*)$  such that
$$P(\mu^*=\mu_-)=\frac{\mu_+-x^*}{\mu_+-\mu_-},\;P(\sg^*=\sg_-)=\frac{\sg_+-y^*}{\sg_+-\sg_-}$$
is the optimal pair.

{\bf Example 1}. Let $H$ be a  constant. i.e. $h_1=0$. It is
evident
$$\min_{(x,y)\in D}\psi(x,y)=\min_{\mu_-\le x\le\mu_+}\frac{h_0^2x^2}{2\mu_Mx-\mu_-\mu_++\sg_+^2}=\min_{0\le t\le1}
\frac{h_0^2(\mu_-+t(\mu_+-\mu_-))^2}{\mu_-^2+t(\mu_+^2-\mu_-^2)+\sg_+^2}.$$
Hence $(x^*,y^*)\in B_{++}$ and we must find
$t_{++}=\arg\min\vp_{++}(t)$. From (\ref{abc}) we have
\begin{equation}
\alpha_{++}=-\frac{\mu_-}{\mu_+-\mu_-}<0,\;
\beta_{++}=-\frac{\mu_{-}^2+\sg_+^2}{\mu_+^2-\mu_-^2},\;
\gamma_{+}=\frac{h_0^2(\Delta\mu)^2}{\mu_+^2-\mu_-^2}=h_0^2\frac{\mu_+-\mu_-}{\mu_++\mu_-}>0.
\end{equation}
Moreover
\begin{equation}
2\beta_{++}-\alpha_{++}=-\frac{\mu_{-}^2-\mu_-\mu_++2\sg_+^2}{\mu_+^2-\mu_-^2}<\frac{\mu_-}{\mu_++\mu_-}<1.
\end{equation}
Thus
\begin{equation}
t_{++}=
  \begin{cases}
    0, & \text{if} \;\; 2\beta_{++}-\al_{++}\le0\\
    \frac{\mu_-\mu_+-\mu_{-}^2-2\sg_+^2}{\mu_+^2-\mu_-^2}, & \text{if}\;\;0<2\beta_{++}-\al_{++}.
  \end{cases}
\end{equation}
Simplifying we obtain
\begin{equation}
t_{++}=
  \begin{cases}
    0, & \text{if} \;\; \mu_+\mu_--2\sg_+^2\le\mu_-^2\\
    \frac{\mu_-\mu_+-\mu_{-}^2-2\sg_+^2}{\mu_+^2-\mu_-^2}, & \text{if}\;\;\mu_-^2<\mu_+\mu_--2\sg_+^2
  \end{cases}
\end{equation}
and
\begin{equation}
\min_{\nu}\frac{(\int_Dh_0\mu\nu(d\mu
d\sg))^2}{\int_D(\mu^2+\sg^2)\nu(d\mu d\sg)}=\vp_{++}(t_{++})=
  \begin{cases}
    \frac{h_0^2\mu_-^2}{\mu_-^2+\sg_+^2}, & \text{if} \;\; \mu_+\mu_--2\sg_+^2\le\mu_-^2\\
    \frac{\mu_+\mu_--\sg_+^2}{\mu_M^2}h_0^2, &
    \text{if}\;\;\mu_-^2<\mu_+\mu_--2\sg_+^2.
  \end{cases}
\end{equation}
By (\ref{opst}) the optimal strategy is
\begin{equation}
\pi^*=h_0\frac{\mu_-+t_{++}(\mu_+-\mu_-)}{\mu_-^2+t_{++}(\mu_+^2-\mu_-^2)+\sg_+^2}=
  \begin{cases}
    \frac{h_0\mu_-}{\mu_-^2+\sg_+^2}, & \text{if} \;\; \mu_+\mu_--2\sg_+^2\le\mu_-^2\\
    \frac{h_0}{\mu_M}, &
    \text{if}\;\;\mu_-^2<\mu_+\mu_--2\sg_+^2.
  \end{cases}
\end{equation}

{\bf Example 2}. Analogously we can consider the case $h_0=0$.
Then $(x^*,y^*)=(\mu_+,\sg_-+t_{-+}(\sg_+-\sg_-))\in B_{-+}$,
\begin{equation}
t_{-+}=
  \begin{cases}
    0, & \text{if} \;\; \sg_+\sg_--2\mu_+^2\le\sg_-^2\\
    \frac{\sg_-\sg_+-\sg_{-}^2-2\mu_+^2}{\mu_+^2-\mu_-^2}, &
    \text{if}\;\;\sg_-^2<\sg_+\sg_--2\mu_+^2,
  \end{cases}
\end{equation}
\begin{equation}
\pi^*=
  \begin{cases}
    \frac{h_1\sg_-}{\sg_-^2+\mu_+^2}, & \text{if} \;\; \sg_+\sg_--2\mu_+^2\le\sg_-^2\\
    \frac{h_1}{\sg_M}, &
    \text{if}\;\;\sg_-^2<\sg_+\sg_--2\mu_+^2.
  \end{cases}
\end{equation}

{\bf Example 3}. let $\mu_-=\mu_+=0$. Then
$\psi(x,y)=\frac{h_1^2y^2}{2\sg_My-\sg_-\sg_+}$ and

$$
t^*=\arg\min_{0\le t\le1}
\frac{h_1^2(\sg_-+t(\sg_+-\sg_-))^2}{\sg_-^2+t(\sg_+^2-\sg_-^2)}=\frac{\sg_-}{\sg_++\sg_-}.
$$
Therefore $\pi^*=\frac{h_1}{\sg_M}$.

\be{rem} The quantity
\begin{equation}
\max_{\nu}\min_\pi F(\pi,\nu).
\end{equation}
is a function of initial capital $x_0$. Minimizing this expression
by $x_0$ we find $x_0^*$ and further construct the optimal
$(\pi^*,\mu^*,\sg^*)$ assuming $h_0=EH-x_0^*$. Therefore we find
 the solution of the problem
\begin{equation}
\min_{x_0,\pi}\max_{(\mu,\sg)}F(\pi,\mu,\sg).
\end{equation}
 \ee{rem}

 \

\section{The alternative approach}

\

For two distinct pairs $(a,b),(c,d)\in\{+,-\}\times\{+,-\}$ such
that $\mu_a^2+\sg_b^2\le\mu_c^2+\sg_d^2$ we define the functions
$f_{abcd}(\pi)=\max(f_{ab}(\pi),f_{cd}(\pi))$, where
$f_{ab}(\pi)=F(\pi,\mu_a,\sg_b)$. Obviously that
$$f_{abcd}(\pi)\le\max_{(a,b)\in\{+,-\}^2}(f_{ab}(\pi))\;\text{and}\;\min_\pi f_{abcd}(\pi)\le\min_\pi\max_{(a,b)\in\{+,-\}^2}(f_{ab}(\pi))$$
Hence by Theorem 3.3 of \cite{dem} (Chapter Vl p.197)
$$\min_\pi\max_{(\mu,\sg)\in D}F(\pi,\mu,\sg)=\min_\pi\max_{(a,b)\in\{+,-\}^2}(f_{ab}(\pi))=\max_{(abcd)}\min_\pi f_{abcd}(\pi).$$
\be{lem} For $\pi_{abcd}=\arg\min_\pi f_{abcd}(\pi)$ we have
\begin{equation}\lbl{abcd}
\pi_{abcd}=
  \begin{cases}
    0, & \text{if}\;\;(h_0\mu_a+h_1\sg_b)(h_0\mu_c+h_1\sg_d)\le0, \\
    {}\\
    \frac{h_0\mu_a+h_1\sg_b}{\mu_a^2+\sg_b^2}, & \text{if}\;\;
    \frac{h_0\mu_a+h_1\sg_b}{\mu_a^2+\sg_b^2}, \frac{h_0\mu_c+h_1\sg_d}{\mu_c^2+\sg_d^2}>
    2\frac{h_0(\mu_c-\mu_a)+h_1(\sg_d-\sg_b)}{{\mu_c^2-\mu_a^2+\sg_d^2-\sg_b^2}}\\
    {}\\
     \frac{h_0\mu_c+h_1\sg_d}{\mu_c^2+\sg_d^2}, & \text{if}\;\;
    \frac{h_0\mu_a+h_1\sg_b}{\mu_a^2+\sg_b^2},
    \frac{h_0\mu_c+h_1\sg_d}{\mu_c^2+\sg_d^2}\le
    2\frac{h_0(\mu_c-\mu_a)+h_1(\sg_d-\sg_b)}{{\mu_c^2-\mu_a^2+\sg_d^2-\sg_b^2}} \\
    {}\\
    2\frac{h_0(\mu_c-\mu_a)+h_1(\sg_d-\sg_b)}{{\mu_c^2-\mu_a^2+\sg_d^2-\sg_b^2}} , &
    \text{if}\;\;\frac{h_0\mu_a+h_1\sg_b}{\mu_a^2+\sg_b^2}
    \le
    2\frac{h_0(\mu_c-\mu_a)+h_1(\sg_d-\sg_b)}{{\mu_c^2-\mu_a^2+\sg_d^2-\sg_b^2}}\le
    \frac{h_0\mu_c+h_1\sg_d}{\mu_c^2+\sg_d^2}\\
    & \;\text{or} \frac{h_0\mu_c+h_1\sg_d}{\mu_c^2+\sg_d^2}
    \le
    2\frac{h_0(\mu_c-\mu_a)+h_1(\sg_d-\sg_b)}{{\mu_c^2-\mu_a^2+\sg_d^2-\sg_b^2}}\le
    \frac{h_0\mu_a+h_1\sg_b}{\mu_a^2+\sg_b^2}.
  \end{cases}
\end{equation}
Moreover
\begin{equation}\lbl{abcd}
f_{abcd}(\pi_{abcd})=
  \begin{cases}
    h_0^2+h_1^2, & \text{if}\;\;(h_0\mu_a+h_1\sg_b)(h_0\mu_c+h_1\sg_d)\le0, \\
    {}\\
    h_0^2+h_1^2-\frac{(h_0\mu_a+h_1\sg_b)^2}{\mu_a^2+\sg_b^2}, & \text{if}\;\;
    \frac{h_0\mu_a+h_1\sg_b}{\mu_a^2+\sg_b^2}, \frac{h_0\mu_c+h_1\sg_d}{\mu_c^2+\sg_d^2}>
    2\frac{h_0(\mu_c-\mu_a)+h_1(\sg_d-\sg_b)}{{\mu_c^2-\mu_a^2+\sg_d^2-\sg_b^2}}\\
    {}\\
    h_0^2+h_1^2-\frac{(h_0\mu_c+h_1\sg_d)^2}{\mu_c^2+\sg_d^2}, & \text{if}\;\;
    \frac{h_0\mu_a+h_1\sg_b}{\mu_a^2+\sg_b^2},
    \frac{h_0\mu_c+h_1\sg_d}{\mu_c^2+\sg_d^2}\le
    2\frac{h_0(\mu_c-\mu_a)+h_1(\sg_d-\sg_b)}{{\mu_c^2-\mu_a^2+\sg_d^2-\sg_b^2}} \\
    {}\\
    h_0^2+h_1^2-\frac{(h_0\mu_a+h_1\sg_b)^2}{\mu_a^2+\sg_b^2}+
    (\mu_a^2+\sg_b^2)\times\\
    (2\frac{h_0(\mu_c-\mu_a)+h_1(\sg_d-\sg_b)}{{\mu_c^2-\mu_a^2+\sg_d^2-\sg_b^2}}-\frac{h_0\mu_a+h_1\sg_b}{\mu_a^2+\sg_b^2})^2,\\
    &
    \text{if}\;\;\frac{h_0\mu_a+h_1\sg_b}{\mu_a^2+\sg_b^2}
    \le2\frac{h_0(\mu_c-\mu_a)+h_1(\sg_d-\sg_b)}{{\mu_c^2-\mu_a^2+\sg_d^2-\sg_b^2}}\\
    &\le\frac{h_0\mu_c+h_1\sg_d}{\mu_c^2+\sg_d^2}\\
    & \;\text{or}\;\; \frac{h_0\mu_c+h_1\sg_d}{\mu_c^2+\sg_d^2}
    \le 2\frac{h_0(\mu_c-\mu_a)+h_1(\sg_d-\sg_b)}{{\mu_c^2-\mu_a^2+\sg_d^2-\sg_b^2}}\\
    &\le\frac{h_0\mu_a+h_1\sg_b}{\mu_a^2+\sg_b^2}
  \end{cases}
\end{equation}
\ee{lem}

{\it Proof}. The minimal value of $f_{ab}(\pi),f_{cd}(\pi)$ are
achieved at $ \frac{h_0\mu_a+h_1\sg_b}{\mu_a^2+\sg_b^2}$ and
$\frac{h_0\mu_c+h_1\sg_d}{\mu_c^2+\sg_d^2}$ respectively.  If
$(h_0\mu_a+h_1\sg_b)(h_0\mu_c+h_1\sg_d)\le0$ then by continuity of
$h_0x+h_1y$ there exists $(x,y)\in D$ such that $h_0x+h_1y=0$ and
$\pi^*=0$. If $(h_0\mu_a+h_1\sg_b)(h_0\mu_c+h_1\sg_d)>0$ then we
assume $h_0\mu_a+h_1\sg_b>0,\;h_0\mu_c+h_1\sg_d>0$.  The roots of
the equation $f_{ab}(\pi)=f_{cd}(\pi)$ are $\pi=0$ and
$\pi=2\frac{h_0(\mu_c-\mu_a)+h_1(\sg_d-\sg_b)}{{\mu_c^2-\mu_a^2+\sg_d^2-\sg_b^2}}$.
There exists three possibilities:

1)$\;\;
    \frac{h_0\mu_a+h_1\sg_b}{\mu_a^2+\sg_b^2}, \frac{h_0\mu_c+h_1\sg_d}{\mu_c^2+\sg_d^2}>
    2\frac{h_0(\mu_c-\mu_a)+h_1(\sg_d-\sg_b)}{{\mu_c^2-\mu_a^2+\sg_d^2-\sg_b^2}},
    $

2)
    $\frac{h_0\mu_a+h_1\sg_b}{\mu_a^2+\sg_b^2},
    \frac{h_0\mu_c+h_1\sg_d}{\mu_c^2+\sg_d^2}\le
    2\frac{h_0(\mu_c-\mu_a)+h_1(\sg_d-\sg_b)}{{\mu_c^2-\mu_a^2+\sg_d^2-\sg_b^2}},$

3)$\frac{h_0\mu_a+h_1\sg_b}{\mu_a^2+\sg_b^2}
    \le
    2\frac{h_0(\mu_c-\mu_a)+h_1(\sg_d-\sg_b)}{{\mu_c^2-\mu_a^2+\sg_d^2-\sg_b^2}}\le
    \frac{h_0\mu_c+h_1\sg_d}{\mu_c^2+\sg_d^2}$

$\text{or} \frac{h_0\mu_c+h_1\sg_d}{\mu_c^2+\sg_d^2}
    \le
    2\frac{h_0(\mu_c-\mu_a)+h_1(\sg_d-\sg_b)}{{\mu_c^2-\mu_a^2+\sg_d^2-\sg_b^2}}\le
    \frac{h_0\mu_a+h_1\sg_b}{\mu_a^2+\sg_b^2}.$

\noindent In each cases the corresponding minimal value calculated
by the equation (\ref{abcd}).

\be{cor} The solution of minmax  problem (\ref{mnmx}) can be given
as
$$
\pi^*=\pi_{a^*b^*c^*d^*},
$$
where $a^*b^*c^*d^*=\arg\max_{(abcd)} f_{abcd}(\pi_{abcd})$
\ee{cor}

 \

\appendix
\section{Appendix}

\

 We need to
find the measure
$\nu_t=t\delta_{(\mu_a,\sg_b)}+(1-t)\delta_{(\mu_c,\sg_d)}$
minimizing the expression
\begin{equation}
\min_{\nu}\frac{(\int_D(h_0\mu+h_1\sg)\nu(d\mu
d\sg))^2}{\int_D(\mu^2+\sg^2)\nu(d\mu d\sg)}
\end{equation}
for $(a,b),(c,d)\in \{-,+\}\times\{-,+\}$. We consider only the
case $0<\mu_-<\mu_+,\;0<\sg_-<\sg_+.$

 Let
 \begin{align}
 \vp(t)=\frac{(\int_D(h_0\mu+h_1\sg)\nu_t(d\mu
d\sg))^2}{\int_D(\mu^2+\sg^2)\nu_t(d\mu d\sg)}\\
\equiv\frac{(h_0\mu_a+h_1\sg_b+t(h_0\Delta\mu+h_1\Delta\sg))^2}
{(\mu_a^2+\sg_b^2+t(\mu_c^2-\mu_a^2+\sg_d^2-\sg_c^2))}.
\end{align}
When $\mu_c^2-\mu_a^2+\sg_d^2-\sg_b^2=0$ and
$h_0\Delta\mu+h_1\Delta\sg=0$, then $\vp(t)=const$.

\noindent If $\mu_c^2-\mu_a^2+\sg_d^2-\sg_b^2=0$ and
$h_0\Delta\mu+h_1\Delta\sg\neq 0$ then
\beq\vp(t)=\frac{1}{\mu_a^2+\sg_b^2}
(h_0\mu_a+h_1\sg_b+t(h_0\Delta\mu+h_1\Delta\sg))^2.\eeq \noindent
 If
$\mu_b^2-\mu_a^2+\sg_d^2-\sg_b^2\neq 0$ and
$h_0\Delta\mu-h_1\Delta\sg\neq 0$ then
\begin{align}
\vp(t)=\gm\frac{(t-\al)^2}{t-\beta}
\end{align}
where $\Delta\mu=\mu_c-\mu_a,\;\Delta\sg=\sg_d-\sg_b$
\begin{equation}\lbl{abc}
\alpha=-\frac{h_0\mu_a+h_1\sg_b}{h_0\Delta\mu+h_1\Delta\sg},\;
\beta=-\frac{\mu_a^2+\sg_b^2}{\mu_c^2-\mu_a^2+\sg_d^2-\sg_b^2},\;
\gamma=\frac{(h_0\Delta\mu+h_1\Delta\sg)^2}{\mu_c^2-\mu_a^2+\sg_d^2-\sg_b^2}
\end{equation}
\be{prop} Let $t^*=\arg\min_{t\in[0,1]}\vp(t)$ and
$\mu_c^2-\mu_a^2+\sg_d^2-\sg_b^2\neq0$ is satisfied. Then for the
case $\gm<0$
\begin{equation}
t^*=
  \begin{cases}
    1, & \text{if}\;\;1\le \alpha\le \beta\;\;\text{or}\;\;1\le 2\beta-\alpha<\beta, \\
    0, & \text{if} \;\; \alpha\le0 \;\;\text{or}\;\;2\beta\le \alpha\\
    \al, & \text{if} \;\; 0<\alpha<1 \\
    2\beta-\alpha, & \text{if}\;\;0<2\beta-\alpha<1
  \end{cases}
\end{equation}
and for the case $\gm>0$
\begin{equation}
t^*=
  \begin{cases}
    1, & \text{if}\;\;1\le \al\;\;\text{or}\;\;1\le 2\beta-\al, \\
    0, & \text{if} \;\; \beta\le \al\le0 \;\;\text{or}\;\;\beta<2\beta-\al\le 0\\
    \al, & \text{if} \;\; 0<\al<1 \\
    2\beta-\al, & \text{if}\;\;0<2\beta-\al<1
  \end{cases}
\end{equation}
\ee{prop} {\it Proof}. Obviously that
\begin{equation}
\vp(t)=\gm\left(t-\al+2(\al-\beta)+\frac{(\al-\beta)^2}{t-\beta}\right)
\end{equation}
and
\begin{equation}\lbl{der}
\vp'(t)=\gm\frac{(t-2\beta+\al)(t-\al)}{(t-\beta)^2}.
\end{equation}
The case $\al=\beta$ is trivial.

I) Let $\gm<0$ and $\al\neq \beta$ then $\beta>1$ and
\begin{align}
\lim_{t\uparrow \beta}\vp(t)=\infty,\;\lim_{t\rightarrow
-\infty}\vp(t)=\infty,\\
\lim_{t\downarrow \beta}\vp(t)=-\infty,\;\lim_{t\rightarrow
\infty}\vp(t)=-\infty
\end{align}
Hence $\vp(t)$ has a minimum on $(-\infty,\beta)$ and has a
maximum on $(\beta,\infty)$. Thus if $\al<\beta$ then as follows
from (\ref{der})  the local minimum is attained at $t=\al$, and if
$\al>\beta$ then $2\beta-\al<\beta$ and the local minimum is
attained at $t=2\beta-\al$.

II) Let $\gm>0$ and $\al\neq \beta$ then $\beta<-1$ and
\begin{align}
\lim_{t\uparrow \beta}\vp(t)=-\infty,\;\lim_{t\rightarrow
-\infty}\vp(t)=-\infty,\\
\lim_{t\downarrow \beta}\vp(t)=\infty,\;\lim_{t\rightarrow
\infty}\vp(t)=\infty
\end{align}
Hence $\vp(t)$ has a maximum  on $(-\infty,\beta)$ and has a
minimum on $(\beta,\infty)$. Thus if $\al>\beta$ then as follows
from (\ref{der}) $t=\al$ is the point of local minimum, and if
$\al<\beta$ then $2\beta-\al>\beta$ and $t=2\beta-\al$ is the
point of local minimum. \qed

Denote by $\vp_{--}(t),\vp_{-+}(t),\vp_{+-}(t),\vp_{++}(t)$ the
function $\vp(t)$ for the cases
$(a,b,c,d)=(-,-,-,+),(+,-,+,+),(-,-,+,-),(-,+,+,+)$ respectively.
We may say that they are functions defined on sides of the
rectangle $D$. Then ({\ref{abc}) takes the form
\begin{equation}\lbl{abcb}
\alpha_{-\pm}=-\frac{h_0\mu_\pm+h_1\sg_-}{h_1\Delta\sg},\;
\beta_{-\pm}=-\frac{\mu_{\pm}^2+\sg_-^2}{\sg_+^2-\sg_-^2},\;
\gamma_{-}=\frac{(h_1\Delta\sg)^2}{\sg_+^2-\sg_-^2}
\end{equation}
\begin{equation}\lbl{abcc}
\alpha_{+\pm}=-\frac{h_0\mu_-+h_1\sg_\pm}{h_0\Delta\mu},\;
\beta_{+\pm}=-\frac{\mu_{-}^2+\sg_\pm^2}{\mu_+^2-\mu_-^2},\;
\gamma_{+}=\frac{(h_0\Delta\mu)^2}{\mu_+^2-\mu_-^2}
\end{equation}
Obviously that $\mu_c^2-\mu_a^2+\sg_d^2-\sg_b^2\neq0$ and $\gm>0$
for this cases. Hence from the previous Proposition we obtain
\be{prop}\lbl{pb2} Let $t_{ab}=\arg\min_{t\in[0,1]}\vp_{ab}(t)$,
for $(a,b)\in\{-,+\}^2$. Then
\begin{equation}
t_{ab}=
  \begin{cases}
    1, & \text{if}\;\;1\le \al_{ab}\;\;\text{or}\;\;1\le 2\beta_{ab}-\al_{ab}, \\
    0, & \text{if} \;\; \beta_{ab}\le \al_{ab}\le0 \;\;\text{or}\;\;\beta_{ab}<2\beta_{ab}-\al_{ab}\le 0\\
    \al_{ab}, & \text{if} \;\; 0<\al_{ab}<1 \\
    2\beta_{ab}-\al_{ab}, & \text{if}\;\;0<2\beta_{ab}-\al_{ab}<1.
  \end{cases}
\end{equation}
\ee{prop}

\end{document}